\documentclass[journal]{IEEEtran}

\usepackage[pdftex]{graphicx}
\usepackage{amsmath,bm}
\usepackage{amssymb}

\ifCLASSINFOpdf
\else
\fi
\begin{document}
%
\title{A Demonstration of Implication Logic Based on Volatile (Diffusive) Memristors}
%
%
%

\author{Yuriy~V.~Pershin,~\IEEEmembership{Senior~Member,~IEEE}
\thanks{Y. V. Pershin is with the Department of Physics and Astronomy, University of South Carolina, Columbia, SC 29208 USA (e-mail: \mbox{pershin@physics.sc.edu}).}
\thanks{Manuscript received June ..., 2018; revised ....}}

\maketitle

\begin{abstract}
 Implication logic gates that are based on volatile memristors are demonstrated experimentally with the use of relay-based volatile memristor emulators of an original design. The fabricated logic circuit involves two volatile memristors and it is capable of performing four fundamental logic functions (two types of material implication and the negations thereof). Moreover, current-voltage characteristics of individual emulators are recorded and self-sustained oscillations in a resistor-volatile memristor circuit are found. The developed emulator offers a great potential for memristive circuit experiments because of its simplicity, similarity of response with volatile memristors, and low cost. Our findings, which are based on emulators, can  easily be reproduced with physical volatile memristors and, thus, open up possibilities for emerging in-memory computing architectures.
\end{abstract}



%
\IEEEpeerreviewmaketitle

\section{Introduction}

%
%
%
%
\IEEEPARstart{D}{uring} the past decade, memristor technology has experienced an explosive growth, which has the potential to revolutionise information processing and storage.
The key advantage of memristors~\cite{chua71a,chua76a} (as well as memcapacitors and meminductors~\cite{diventra09a}) over the traditional electronic components is the possibility to store and process information on the same physical location. Memristive behavior has been observed in many systems and devices~\cite{Pershin11a}. Up to now, however, most attention has been focused on devices with non-volatile storage capability~\cite{PAN20141}. The future applications of non-traditional memristors are still not fully understood, and their theoretical and circuit-level models are still at an early stage of development.

The present paper explores an in-memory computing application of volatile memristors, namely,  memristors capable of storing information only when connected to a power source. Specifically, we will limit ourselves to devices exhibiting two possible resistance states (ON and OFF states) in a finite range of voltages and switching to the OFF state when a smaller voltage is applied. Several physical systems satisfy these requirements, including NEMS switches~\cite{Nieminen02,Sun16}, Mott memristors~\cite{pickett2013scalable}, graphene field emitters~\cite{Kleshch15}, and diffusive memristors~\cite{wang2017memristors}. The last system has recently attracted attention because of its promising characteristics for the use in artificial neural networks~\cite{wang2017memristors}, random signal generators~\cite{jiang2017novel}, and sensing applications~\cite{yoon2018artificial}. Physically, in such diffusive memristors Ag atoms
spread under electrical bias and regroup spontaneously under zero/small bias because of interfacial energy minimization~\cite{wang2017memristors,jiang2017novel}.
Moreover, it was shown that two Mott memristors can be used to build a neuristor~\cite{crane1960neuristor}, which is an electronic analog of the Hodgkin-Huxley axon. In what follows we will keep our discussion general, without referring to any particular physical realisation of volatile memristors. In our electronic circuit experiments, the volatile memristors are represented by  emulators built out of conventional electronics components (resistors and relays). The volatile memristor emulator is developed as a part of the present work.

This paper experimentally demonstrates the implication logic~\cite{borghetti10a} gates based on volatile memristors. Previously, in-memory logic gates were realised experimentally by employing bipolar non-volatile memristors~\cite{borghetti10a}, and explored theoretically based on  bipolar non-volatile memristors (see, e.g., \cite{Kang12a,Kvatinsky14b,Sirakoulis14a,Zhang15a,Swartzlander17a}), unipolar non-volatile memristors~\cite{Sun11a,amrani2016logic}, memcapacitors~\cite{traversa14a,pershin15a}, and volatile memristors (graphene field emitters~\cite{Pershin17a}). The advantage of using memory devices in logic circuits is that they can serve simultaneously as a gate and latch. Here, we employ volatile memristors to create a polymorphic implication logic circuit and we demonstrate four kinds of fundamental logic gates using the same circuit. To the best of our knowledge, this is the first experimental realisation of implication logic gates based on volatile memristors. Moreover, the response of individual emulators is explored. It is found that a simple resistor-volatile memristor circuit can exhibit self-sustained oscillations with a pattern involving both regular and random components.

The rest of this paper is organised as follows. Section \ref{sec:2a} introduces the relay-based emulator of volatile memristors and provides details on its specific realisation and response. Self-sustained oscillations in the resistor-volatile memristor circuit are briefly considered in Section \ref{sec:2b}. Section \ref{sec:3} presents the implication logic gates based on volatile memristors. One of our main results is the experimental demonstration of four fundamental logic functions using the same circuit, which is contained in Section \ref{sec:3}. Our concluding remarks are given in Section \ref{sec:4}.

\section{Volatile Memristor Emulator} \label{sec:2}

\subsection{Emulator} \label{sec:2a}

Memristor emulators~\cite{biolek2014memristor} are valuable tools for circuit prototyping when physical memristors are not accessible. A number of emulator designs are available in the literature based either on analog~\cite{Muthuswamy10a,kim2012memristor,Yu2014a} or digital~\cite{pershin09c,kolka2012hybrid} techniques. With rare exceptions~\cite{Asapu15a,ascoli2016class}, a common feature of memristor emulators  is the need for an external power source. Here, we show that the volatile memristors can be emulated in a very simple way and at low cost. The proposed emulator operates without an external power source and it demonstrates a high similarity to the response of volatile memristors.

\begin{figure}[t]
(a) \centering \includegraphics[width=0.8\columnwidth]{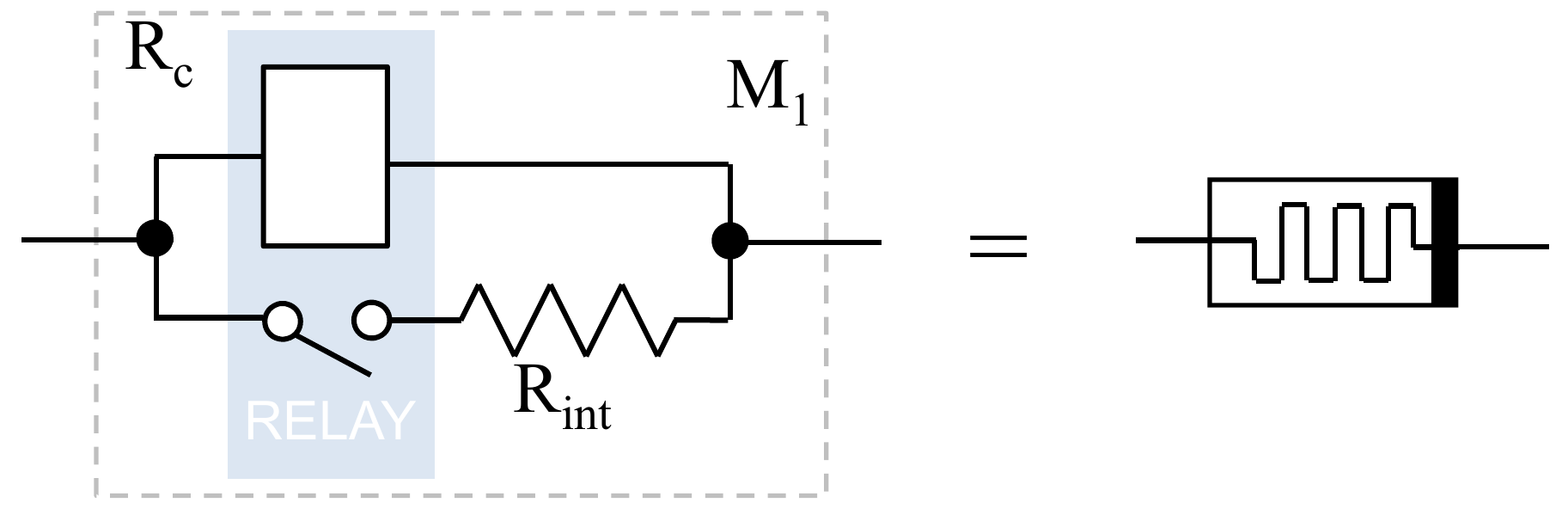} \\
\vspace{0.5cm}
(b) \centering \includegraphics[width=0.7\columnwidth]{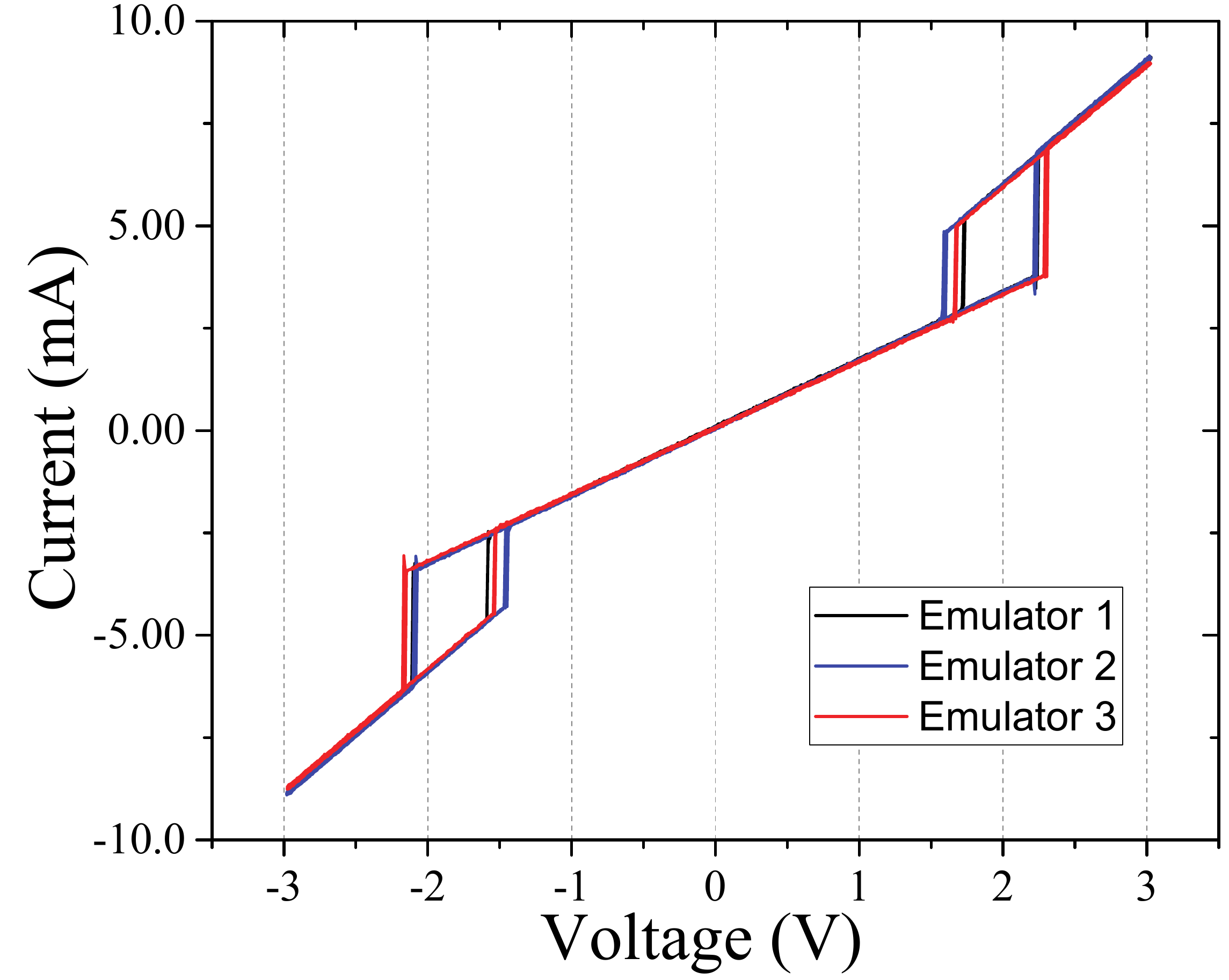}
\caption{(a) Schematics of volatile memristor emulator. An effective two-terminal volatile memristive system
is formed by connecting the relay coil in parallel with series-connected resistor and reed switch. (b) Current-voltage characteristics of three physically different emulators with $R_{int}=680$ $\Omega$.}
\label{fig:1}
\end{figure}

Fig. \ref{fig:1}(a) shows the emulator schematics. The volatile memristor emulator consists of a reed relay and resistor, forming an effective two-terminal memristive system. At lower applied voltages, the reed switch is open. In this case, the emulator resistance equals the coil resistance $R_{OFF}=R_c$. Meanwhile, at higher applied voltages, the switch is closed and the total resistance is $R_{ON}=R_cR_{int}/(R_c+R_{int})$. The interval between the pull-in and drop-out  voltages of relay is the bistability (memory) region.


Three identical volatile memristor emulators were created and their current-voltage characteristics were measured.
In the present experiments, reed relays with the coil resistance of $R_c=600$ $\Omega$ and nominal operating voltage of 5 V are employed (part number HI05-1A66, Standex-Meder Electronics). Fig. \ref{fig:1}(b) shows that the current-voltage characteristics of different emulators are very close to each other. According to Fig. \ref{fig:1}(b), at positive voltages, the OFF to ON transition occurs at $V_{th}\approx 2.2$ V, while the ON to OFF transition takes place at $V_{hold}\approx 1.6$ V. Moreover, the hysteresis region in the negative domain is slightly shifted to lower voltage amplitudes, which is likely due to an asymmetry in the reed switch response with respect to the magnetic field direction.

In addition, mention should be made of the inductive effects originating from
the relay coil. The relay coil can be represented by a resistor and inductor connected in series and described by the impedance of the form $Z=\sqrt{R^2_{int}+\left( \omega L \right)^2}$, where $\omega$ is the angular frequency of input and $L$ is the coil inductance. It follows from this expression that the resistive response is dominant at lower frequencies and transforms into an inductive response at higher frequencies. The transition frequency $\nu_t$ can be estimated from the condition of equal contributions of the resistive and inductive components to the impedance, namely: $R_{int}=2\pi \nu_t L$. In the present realisation of the emulator, the coil inductance $L= 0.17$~H leads to $\nu_t=560$~Hz. The inductive effects should be considered when designing circuits with relay-based emulators operating at higher frequencies.

\subsection{Resistor-volatile Memristor Circuit} \label{sec:2b}

\begin{figure}[t]
\centering (a)  \includegraphics[width=0.65\columnwidth]{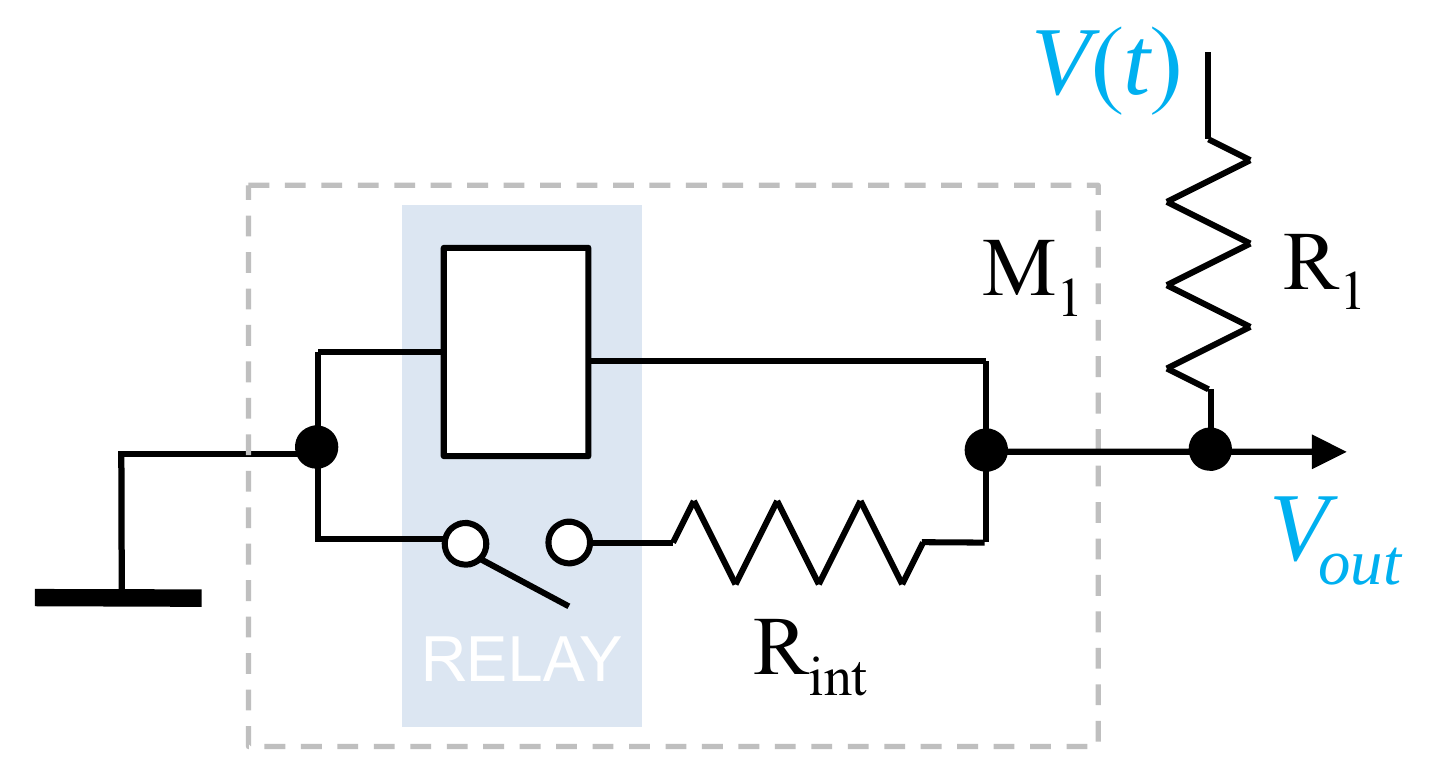}\\
\vspace{0.5cm}
\centering (b) \includegraphics[width=0.9\columnwidth]{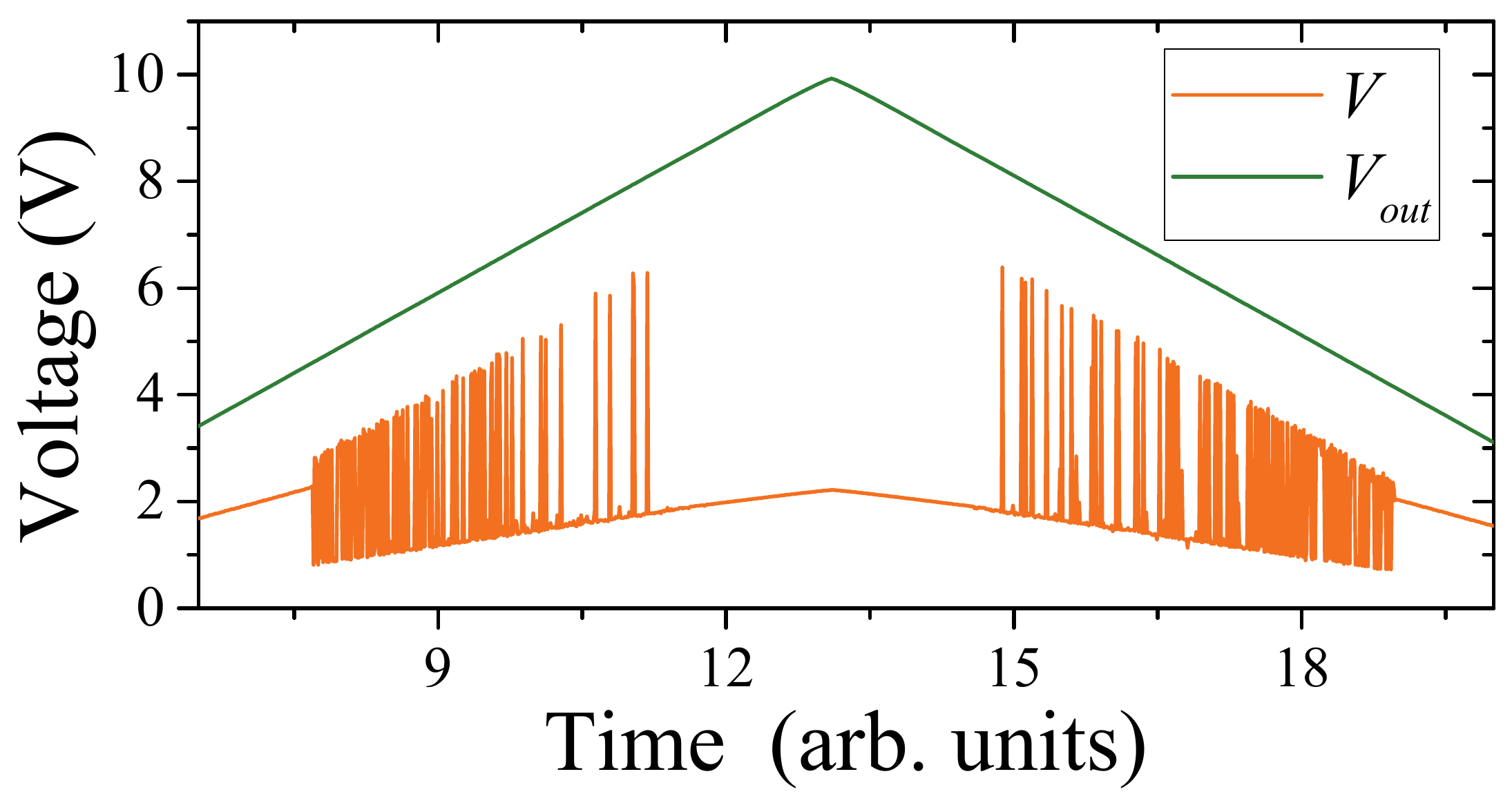}\\
\vspace{0.5cm}
\centering (c) \includegraphics[width=0.75\columnwidth]{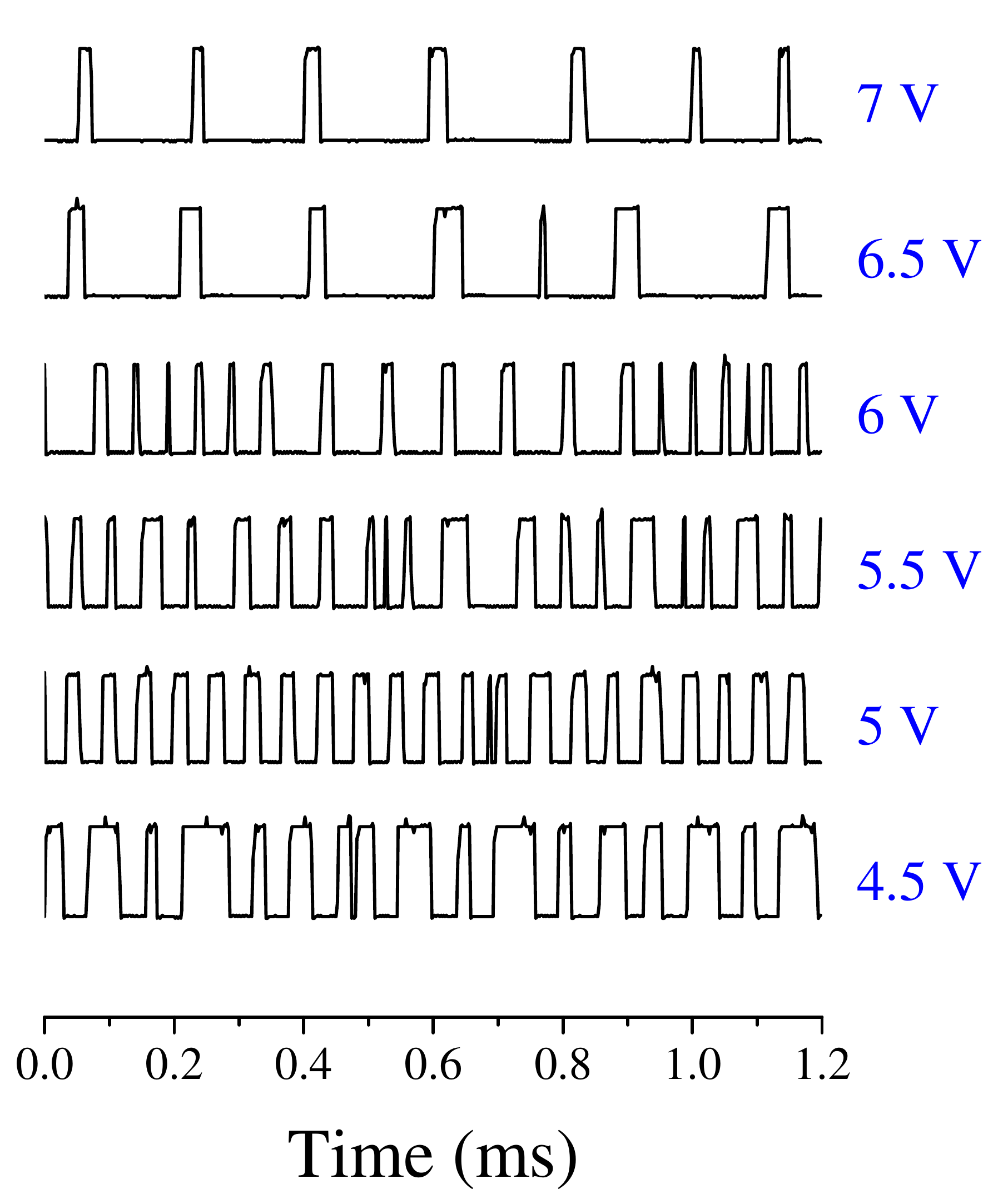}

\caption{(a) Resistor-volatile memristor circuit. (b) Applied and output voltages as functions of time in the resistor-volatile memristor circuit with $R_{int}=220$ $\Omega$ and $R_1=680$ $\Omega$. (c) Digitised $V_{out}$ measured
at the output of a comparator at several constant values of the applied voltage $V$ (indicated on the plot). The curves are shifted for clarity.
}
\label{fig:2}
\end{figure}

To better understand the emulator behavior in electronic circuits, consider a circuit of series-connected resistor and memristor subjected to an applied voltage $V(t)$ (see Fig. \ref{fig:2}(a)). An interesting (and potentially useful) feature of this circuit is the possibility of self-sustained memristance (memory resistance~\cite{chua76a}) oscillations. These oscillations are clearly seen in Fig. \ref{fig:2}(b) showing the response of resistor-volatile memristor circuit to the applied voltage of sawtooth wave form. Technically speaking, the oscillations occur at such applied voltages when in the $R_{ON}$ state of memristor $V_M>V_{th}$ and in the $R_{OFF}$ state $V_M<V_{hold}$. Under these conditions, the memristor will continuously switch back and forth between its low- and high-resistance states. In fact, the same oscillation mechanism works in systems with negative differential resistance.

To derive the necessary condition for the oscillations, consider the resistor-volatile memristor circuit at the onset of switching, namely: assuming that the voltage across M$_1$ is $V_M=V_{th}$ and $R_M=R_{OFF}$. In this case, the applied voltage $\tilde{V}$ is given by
\begin{equation}\label{eq:1}
\tilde{V}=\frac{R_1+R_{OFF}}{R_{OFF}} V_{th}.
\end{equation}
The switching into $R_{ON}$  drops $V_M$ to
\begin{equation}\label{eq:2}
V'_M=\frac{R_{ON}}{R_1+R_{ON}} \tilde{V}.
\end{equation}
If, after this switching, $V'_M<V_{hold}$ then the memristor will switch back into the $R_{OFF}$ state, and so on. In other words, the resistor-volatile memristor circuit will exhibit self-sustained oscillations. By combining the inequality $V'_M<V_{hold}$ with Eqs. (\ref{eq:1}), (\ref{eq:2}) one finds
\begin{equation}\label{eq:3}
\frac{R_{ON}(R_1+R_{OFF})}{R_{OFF}(R_1+R_{ON})}  V_{th}<V_{hold},
\end{equation}
which is the necessary condition for the existence of circuit instability. Clearly, the circuit is stable in the limit of $R_{ON}\rightarrow R_{OFF}$ (note that $V_{th}>V_{hold}$) and unstable at some smaller values of $R_{ON}$.


In the measurements, the signal from the resistor-volatile memristor circuit was transformed to the standard 0 V-(+5 V) logic levels using a comparator with the threshold voltage set at about $2.5$~V. Fig. \ref{fig:2}(c) presents examples of comparator output for several
constant values of the applied voltage $V$. This plot demonstrates that both the frequency and probability of logic "1" in the output signal depend on $V$. The output signal contains the regular (most clearly seen at $V=5$ V curve) and random components, as well as a combination of  frequencies ($V=6$ V curve). From the physics point of view, the random component can be associated with probabilistic sticking/unsticking of relay reeds and/or their complex dynamics under Fig. \ref{fig:2}(a) circuit conditions.

\section{Implication Logic Gates} \label{sec:3}

\begin{figure}[t]
\centering (a)\;\;\;\includegraphics[width=0.7\columnwidth]{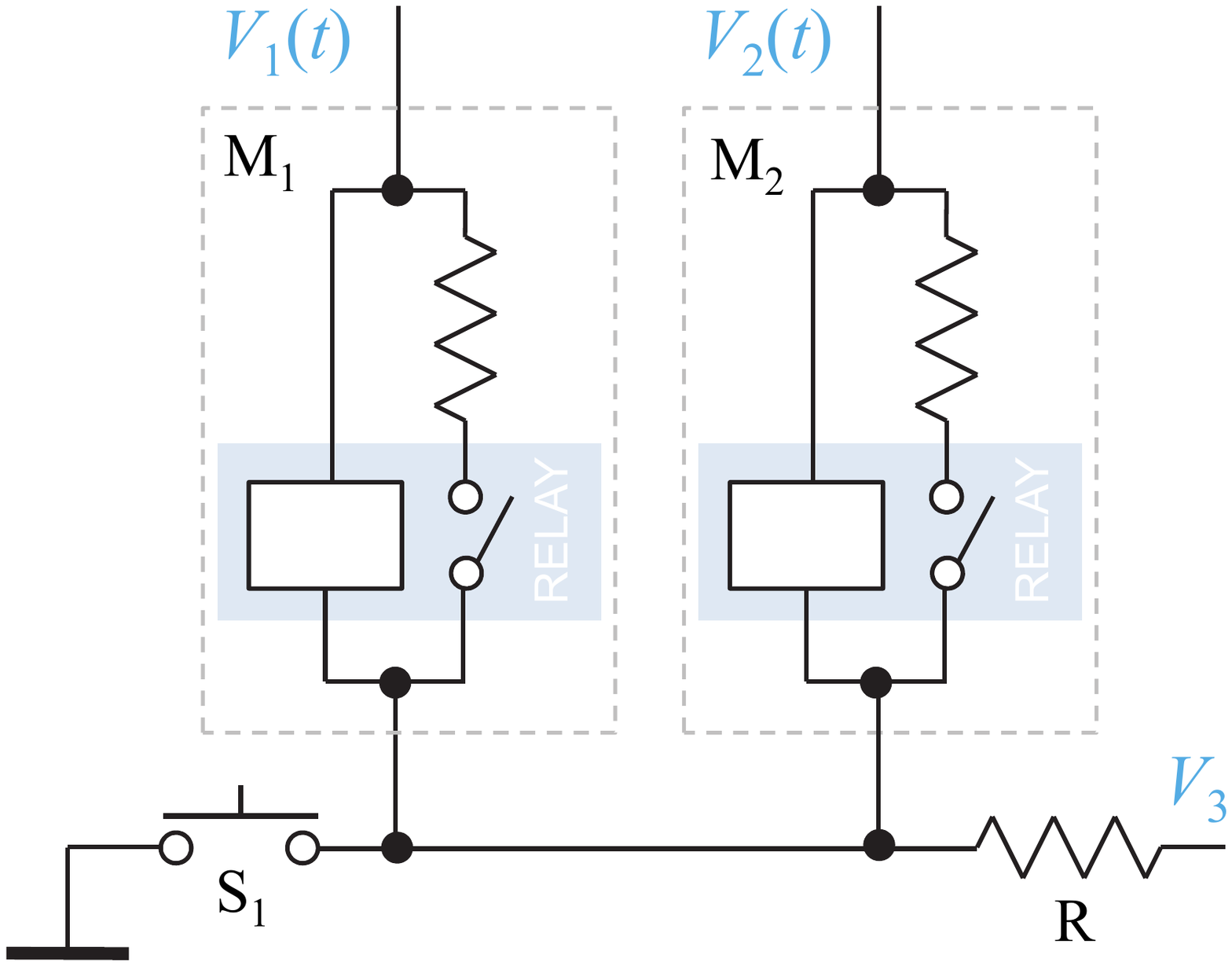} \\
\vspace{3mm}
\centering (b)\;\;\;\includegraphics[width=0.65\columnwidth]{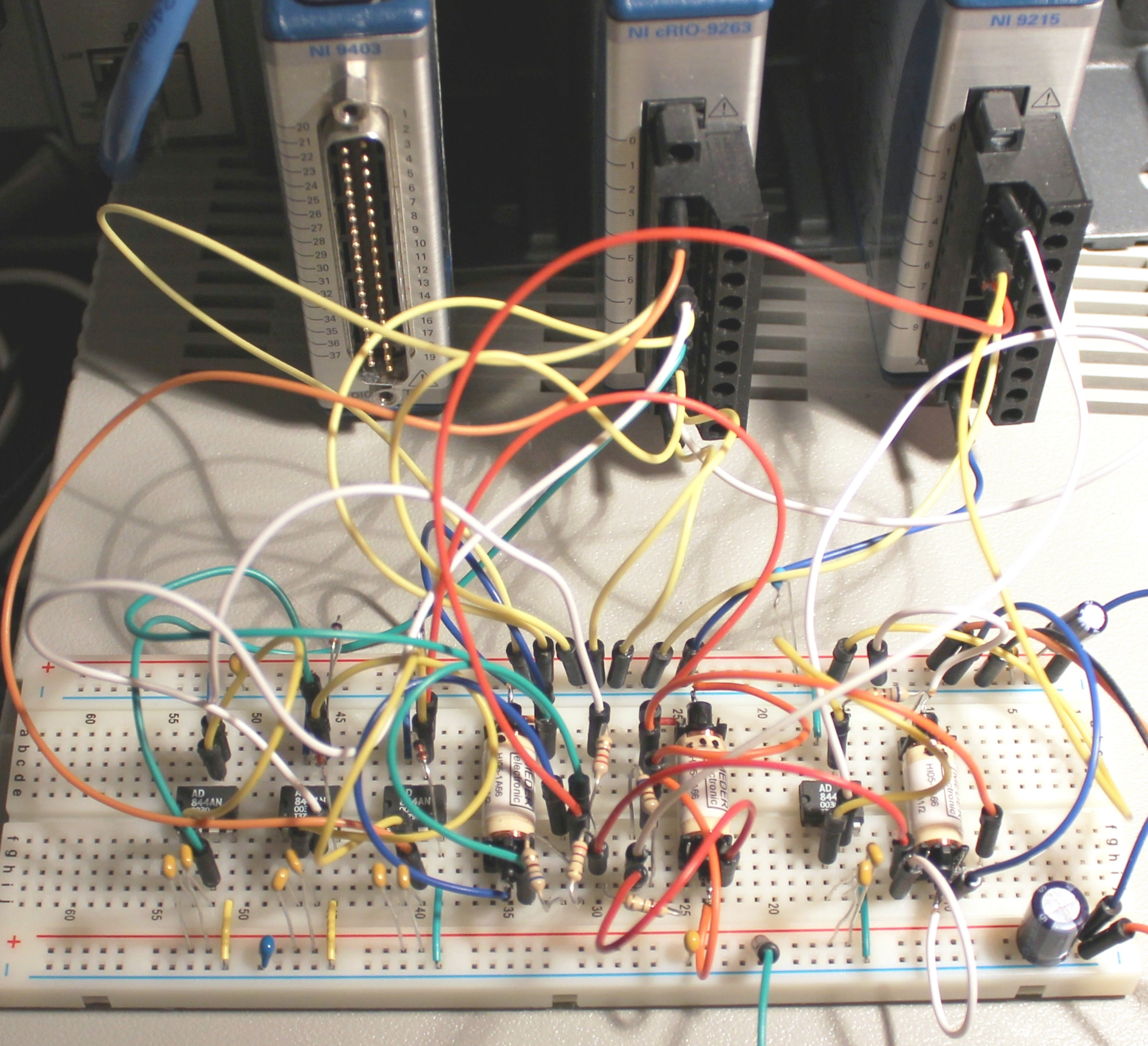}
\caption{(a) Implication logic circuit. In the present measurements, the switch S$_1$ is implemented by a relay. (b) Photograph of experimental setup. Two memristor emulators are located in the center, while the switch S$_1$ is to the right. Operational amplifiers are used as buffers.
}
\label{fig:3}
\end{figure}

The implication logic circuit that is considered in this work is slightly different from the circuit based on non-volatile memristors~\cite{borghetti10a}. Modifications are needed to ensure that the volatile memristors stay in their bistable (hysteresis) regions between the operations. The selected circuit design and its experimental realization are presented in Fig. \ref{fig:3}. In particular, Fig. \ref{fig:3}(a) shows a circuit comprising two volatile memristors, resistor, and switch. Three voltage sources are used to drive the circuit. It is convenient to split the calculation sequence into three phases: initialisation, hold, and calculation. The switch is closed in the initialisation and hold phases, and it is opened within the calculation phase to induce a gate operation.

\begin{figure}[t]
\centering (a) \includegraphics[width=0.78\columnwidth]{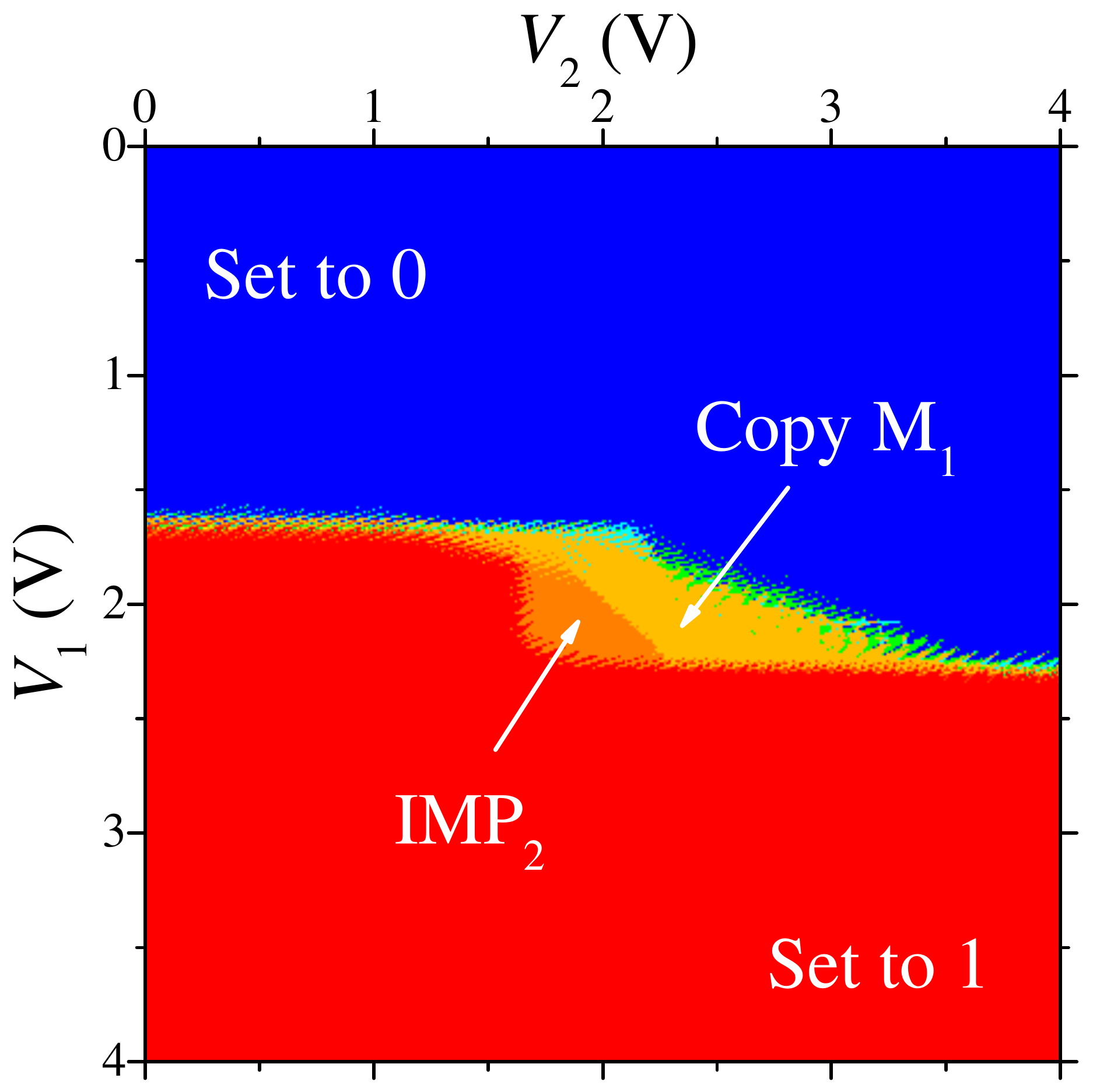} \\
\centering (b) \includegraphics[width=0.78\columnwidth]{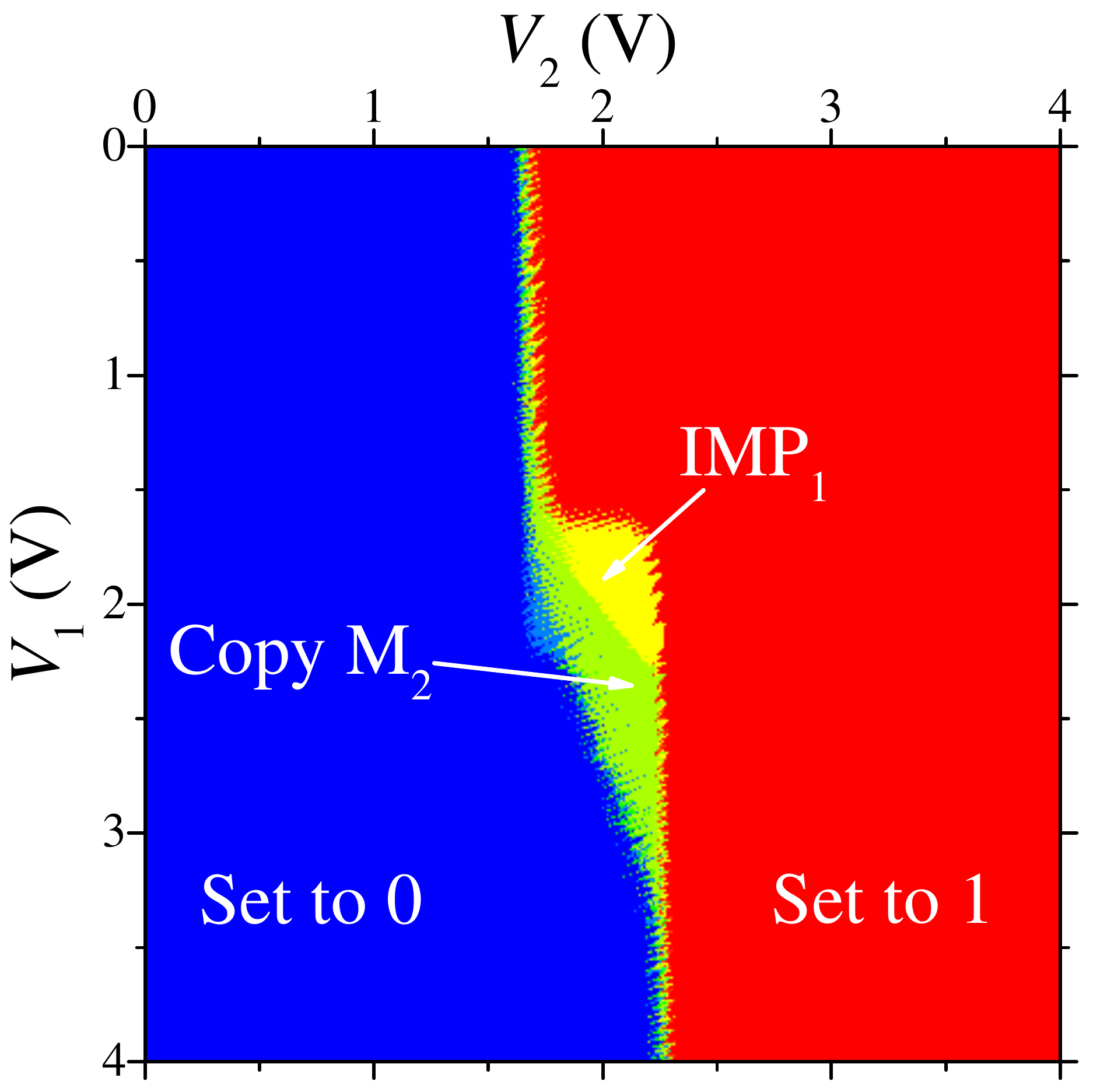}
\caption{Logic gate type as a function of voltage amplitudes $V_1$ and $V_2$ (in the calculation phase). Plot (a) shows the gate type related to the final state of M$_1$. Plot (b) presents the gate type related to the final state of M$_2$. The measurements were performed at $V_3=-1.9$~V. Here, IMP$_{1}$ denotes M$_{1}\rightarrow $M$_{2}$, and IMP$_{2}$ denotes M$_{2}\rightarrow $M$_{1}$.}
\label{fig:4}
\end{figure}

To store the information, S$_1$ is kept closed and $V_{0}=1.9$~V is applied to M$_1$ and M$_2$ ($V_1=V_2=V_{0}$). To initialise the memristor state, 0 V or 5 V is applied to a given memristor. The calculation is performed from the hold phase by changing $V_1$ and $V_2$ to desired values, setting $V_3$, opening and closing the switch (the calculation phase). The {\it entire calculation sequence} consists of the following steps: initialisation, hold, calculation, hold. Note that the calculation results are stored in the final states of memristors. The final states of the memristors were monitored using small resistors connected in series with memristors and measuring the voltage drops across these resistors. To eliminate the effect of self-sustained oscillations discussed in Sec. \ref{sec:2b} (or at least to reduce it to some insignificant areas in the parameters space), emulators with relatively large $R_{int}=680$ $\Omega$ were used in combination with a smaller common resistor ($R=220$  $\Omega$).

Diagrams of logic operations are obtained following the approach introduced in Ref.~\cite{pershin15a}. For each pair of input combinations ((0,0), (0,1) (1,0), (1,1)), the final states of memristors are measured after applying the entire calculation sequence. The type of logic operation is identified using a code~\cite{pershin15a,Pershin17a} calculated based on the final states of memristors. The code is an integer number (from 0 to 15) that encodes the logic operation type and it is calculated as described in Ref.~\cite{pershin15a}. The code is translated to the logic operation type with the help of the table from Ref.~\cite{pershin15a}. Overall, our measurements confirm the possibility of logic operations previously predicted theoretically for the case of graphene autoemission memristors~\cite{Pershin17a}.

Fig. \ref{fig:4} presents measurement results based on Fig. \ref{fig:3} circuit taken at $V_3=-1.9$~V. The logic operation type is plotted as a function of $V_1$ and $V_2$ applied in the calculation phase. Fig. \ref{fig:4} indicates that two types of material implication, M$_{1}\rightarrow $M$_{2}$ and M$_{2}\rightarrow $M$_{1}$, can be realised by Fig. \ref{fig:3} circuit at $V_3=-1.9$~V. Moreover, note that Fig. \ref{fig:4}(a) and Fig. \ref{fig:4}(b)
can be transformed to each other under the flip across the $V_1=V_2$ diagonal and interchange of indices 1 and 2 in the operation type (this property stems from the symmetric connections of M$_1$ and M$_2$ in the circuit).

\begin{figure}[t]
\centering \includegraphics[width=0.78\columnwidth]{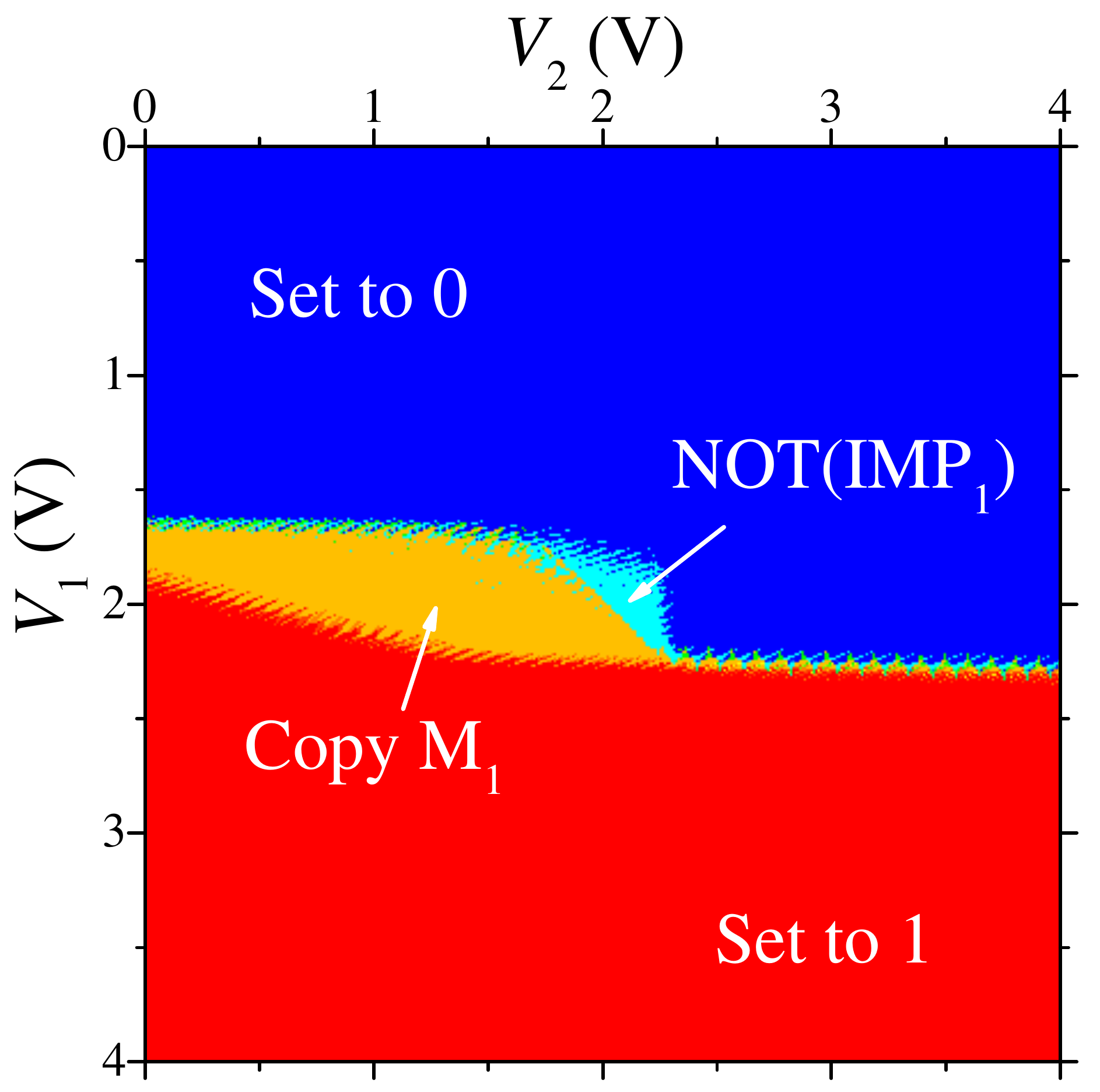}
\caption{Logic gate type as a function of voltage amplitudes $V_1$ and $V_2$. This plot is generated based on the final result stored in M$_1$. The measurements were performed at $V_3=-1.2$~V. Here, NOT(IMP) is the negation of implication.}
\label{fig:5}
\end{figure}

The circuit functionality is changed when $V_3$ is shifted to -1.2~V. The map of logic functions for this case is presented in Fig. \ref{fig:5} for the result stored in M$_1$. Fig. \ref{fig:5} shows that there is a region of voltages in which the negation of implication NOT(IMP$_1$) is realised. Symmetrically, under appropriate conditions, the final state of M$_2$ stores the result of another type of the negation of implication (NOT(IMP$_2$)).

\section{Discussion and Conclusion} \label{sec:4}

In conclusion, the possibility of in-memory computing based on volatile (diffusive) memristors has been demonstrated. Using two volatile memristor emulators, the implication logic circuit was created and four kinds of fundamental logic gates were shown experimentally.  This type of operation is fundamentally different from the case of traditional logic gates having a predetermined functionality. Specifically, in addition to the trivial operations (set to 1, set to 0 and copy the initial states), the following fundamental~\cite{whitehead1912principia} logic gates have been demonstrated: IMP$_1$, IMP$_2$, NOT(IMP$_1$), and NOT(IMP$_2$).
Moreover, self-sustained oscillations were measured in the resistor-volatile memristor circuit. It has been  found that the voltage oscillations involve both regular and random components, which shows the potential application of the resistor-volatile memristor circuit in the area of random number generation.

The present work broadens the opportunities for exploiting volatile memristors for information processing and storage. Compared to the traditional implication logic circuits that are based on non-volatile memristors~\cite{borghetti10a}, volatile memristive circuits are slightly more complex because volatile devices require a power source to store information. We note that several non-idealities of real memristors such as the stochastic component in their dynamics~\cite{jiang2017novel} and variability of device parameters were not captured by our emulators. Such non-idealities (which do also exist in non-volatile memristors) must be considered when designing future in-memory computing circuits/systems.

The volatile memristor emulator that was introduced in this work offers a low cost and simple design alternative to physical memristors for the use in rapid memristive circuit prototyping. Conceptually, its operation
principles can help to explain the operation of volatile memristors, as well as of
circuits based thereof. It is anticipated that the volatile memristor emulators may also find use in undergraduate teaching laboratories to teach  memristor technology and provide relevant hands-on experience.


\section*{Acknowledgement}

The author would like to thank V. A. Slipko for our useful discussions.

\ifCLASSOPTIONcaptionsoff
  \newpage
\fi

\bibliographystyle{IEEEtran}
\bibliography{memcapacitor1}

\end{document}